\documentclass[journal]{IEEEtran}

\ifCLASSINFOpdf
\else
   \usepackage[dvips]{graphicx}
\fi
\usepackage{url}
\usepackage{graphicx}
\usepackage{amsmath}

\begin{document}

\title{Acoustic Echo Control Based on \\Sound Object Identification
for Suppressing \\
Howling Caused by Complicated Acoustic Paths}

\author{Osamu Hoshuyama, \IEEEmembership{Senior Member, IEEE}
\thanks{O. Hoshuyama is with Future Design Research Laboratory, Kyocera Corporation, Japan
(e-mail: osamu.hoshuyama.nf@kyocera.jp).}}

\markboth{Submitted to IEEE Signal Processing Letters, Aug. 17, 2026}
{Hoshuyama: Acoustic Echo Control Based on Sound Object Identification}
\maketitle

\begin{abstract}
This paper proposes 
acoustic echo control based on sound object identification
for suppressing acoustic echo and howling in conferencing environments
with complicated acoustic paths,
where multiple hands-free terminals coexist in the same room.
Conventional acoustic echo cancellers target fixed intra-device echo paths;
however, unintended paths, for example, those formed via inter-terminal communication,
are difficult to control and can lead to howling.
Instead of estimating echo paths, the proposed approach identifies sound objects
and keeps channels muted by default, allowing pass/playback
only when the signal is judged not identical to recently observed objects.
This breaks echo loops caused by repeated reproduction of the same sound object
and can be viewed as an extension of classical voice switching toward conditional half-duplex operation.
Technical challenges and connections to related techniques are discussed, and
a simulation shows howling suppression together with a trade-off against speech quality.
\end{abstract}

\begin{IEEEkeywords}
Echo Cancellers, Telecommunication, Acoustic Communication, Speech Enhancement.
\end{IEEEkeywords}

\IEEEpeerreviewmaketitle

\section{Introduction}
Hands-free terminals such as speakerphones are now routinely used for voice calls.
In hands-free communication, acoustic echo and howling due to acoustic feedback
from the loudspeaker to the microphone are suppressed by an acoustic echo canceller (AEC)
\cite{Sondhi1967,Bennesty2001,Enzner2014,Haensler2005,NeuralKalman2024}.
An AEC estimates the acoustic feedback path with an adaptive filter and cancels the echo.
However, when multiple hands-free terminals operate in the same room,
unexpected echo paths can arise through a communication server between terminals,
and conventional AECs cannot suppress them.
Manual mute operations intended to prevent this are unreliable;
if several terminals are unmuted at once, echo and howling readily occur.
In virtual and metaverse conversations, and with the spread of hearables,
intentionally muting devices may become even harder, so solving echo and howling remains important.
Proximity-aware multi-device control has begun to appear in practice~\cite{Marini2024},
but it still depends on restricted sensing and inter-device coordination.

This paper proposes acoustic echo control that identifies sound objects
and controls their playback, rather than estimating echo paths.
Challenges toward realization are discussed, and a simulation shows that howling can be suppressed.

\section{Acoustic Echo and Howling with Multiple Terminals}
\begin{figure}[tb]
\begin{center}
\includegraphics[width=\columnwidth]{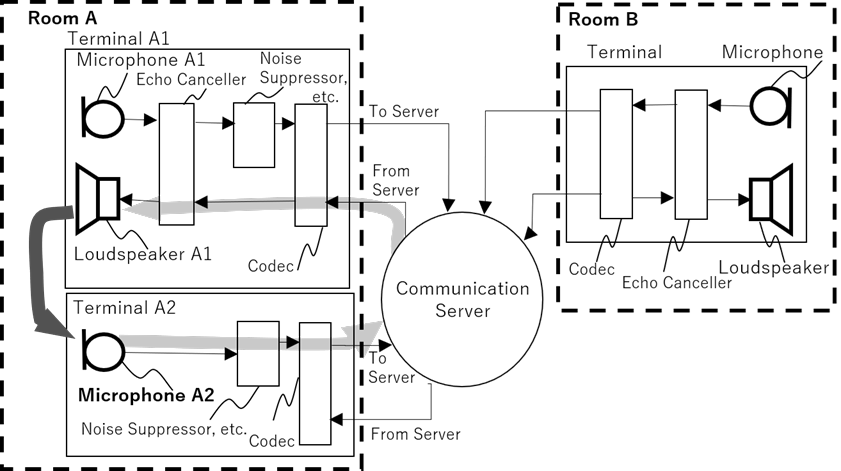}
\caption{Echo paths with multiple hands-free terminals in the same room.
Each terminal consists of a microphone, AEC, noise suppressor, codec, and loudspeaker,
interconnected via a communication server.
Leakage from the A2 headphones is neglected.
A signal from microphone A2 can be played back from loudspeaker A1 through the server
and return to microphone A2, forming a howling loop.}
\label{EcSystemNw01}
\end{center}
\end{figure}

As a typical difficult case, consider the setup in Fig.~\ref{EcSystemNw01}:
one speakerphone-type hands-free terminal A1 and one headset-type terminal A2
placed a short distance apart in the same room.
Leakage from the A2 headphones to its microphone is neglected.
Terminal A1 includes an AEC (with an echo suppressor), so direct intra-device feedback is removed.
The terminals also apply nonlinear processing such as noise suppression (NS) and dynamic range control,
and loudspeaker A1 (including its amplifier) exhibits saturation.

Consider the case where the microphones of A1 and A2 are unmuted simultaneously,
e.g., to send the A2 talker's voice clearly to another room (Room~B in the figure).
Sound entering microphone A2 is encoded, sent through the communication server,
and played from loudspeaker A1.
That playback re-enters microphone A2, forming acoustic echo that can grow into howling.

Usually, unmuting microphone A2 is disallowed, or loudspeaker A1 must be muted first.
Such mute discipline breaks conversational flow and is a major reason why
voice conferences feel inferior to face-to-face meetings.
With hearables, users may join an ongoing call while moving,
creating unexpected echo paths.

Conventional path-estimation AECs cannot realistically handle these inter-terminal paths,
which include the network, nonlinear in-device processing, delay variation,
and user mute actions.
Even with deep learning, solving the problem by path estimation remains extremely difficult.

\section{Proposed Approach}
The proposed approach is called {\em sound-object-based echo control}.
It does not estimate an acoustic path and subtract echo.
Instead, it identifies whether sound objects such as utterances or room sounds
reappear, and controls pass/playback accordingly.
Sound objects are taken to be speech segments of tens to hundreds of milliseconds.

\subsection{Basic Policy}
To prevent howling, channels are muted by default and unmuted for pass/playback
only when a signal is judged {\em not identical} to sound objects recently observed
at the same terminal.
Because only sound objects are used, without path information,
echo loops can in principle be broken even for unintended paths that include
network delay and nonlinear in-device processing.
This control can be viewed as an extension of classical voice-switched half-duplex
operation~\cite{Busala1960,Haensler1992} with sound object identification:
not permanent half-duplex, but conditional gating that closes locally when identity is detected.

\begin{figure*}[tb]
\begin{center}
\includegraphics[width=\textwidth]{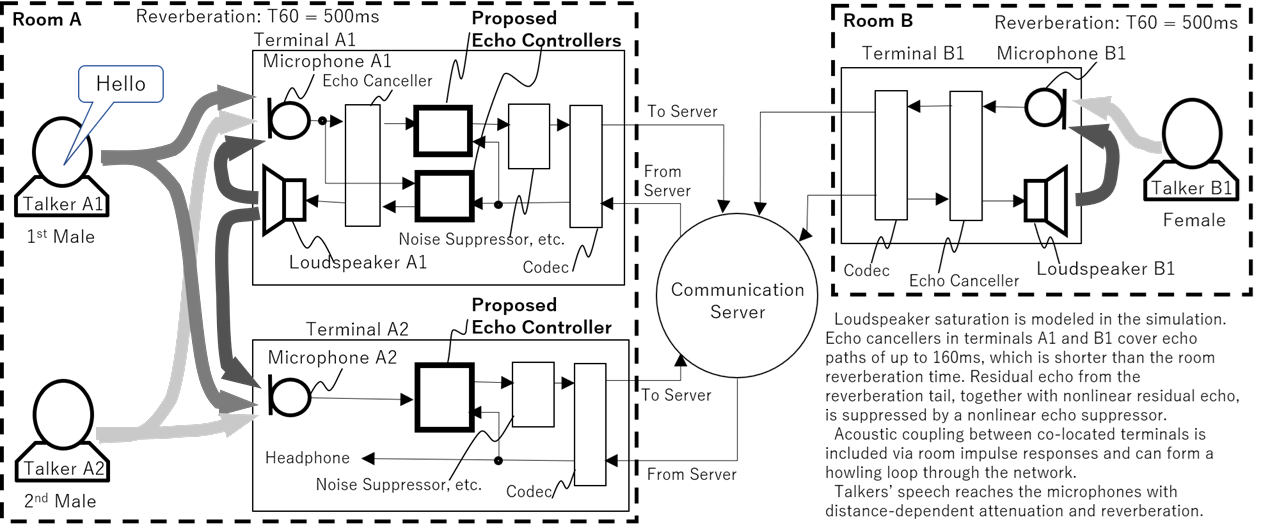}
\caption{Proposed echo control and simulation setup.
Thick-lined blocks denote the proposed control.
On both the microphone and loudspeaker sides, a signal is passed only when
sound-object identification indicates that it is likely different from stored objects.
The simulation uses two rooms and three terminals:
hands-free A1 and microphone-only A2 in Room~A (one male talker near each),
and hands-free B1 in Room~B (one female talker).
Both rooms have $T_{60}=500$\,ms.}
\label{EcSystemNwProp}
\end{center}
\end{figure*}

Figure~\ref{EcSystemNwProp} shows a terminal implementation.
On the transmit side, a signal is sent to the server only when identification
against stored receive-side objects indicates that it is unlikely to stem from the same utterance.
On the receive side, playback is allowed only when the receive object is judged different
from microphone-side objects.
Duplicate reproduction paths of the same object are thereby blocked in principle.

\subsection{Processing Blocks}
Processing comprises three elements.
End-to-end joint processing may be possible later; here the elements are separated
for discussion, challenge analysis, and verification.

\begin{enumerate}
\item {\bf Extraction and buffering of sound objects.}
Extract objects from the microphone or receive signal and retain them
for a duration on the order of the network delay and room reverberation.
\item {\bf Sound object identification.}
Compute a similarity or identity probability between the current
transmit/receive object and the buffered set.
\item {\bf Playback control.}
Control gain from the identification result:
mute when identity is likely; pass/playback otherwise.
If comparison candidates are insufficient, keep mute (safe-side).
\end{enumerate}

Even if identification errs, a non-persistent error tends to cause only a brief echo
rather than sustained howling.
Over-muting desired speech, however, reduces intelligibility,
so the trade-off between howling suppression and speech quality is fundamental.

\begin{figure*}[tb]
\begin{center}
\includegraphics[width=\textwidth]{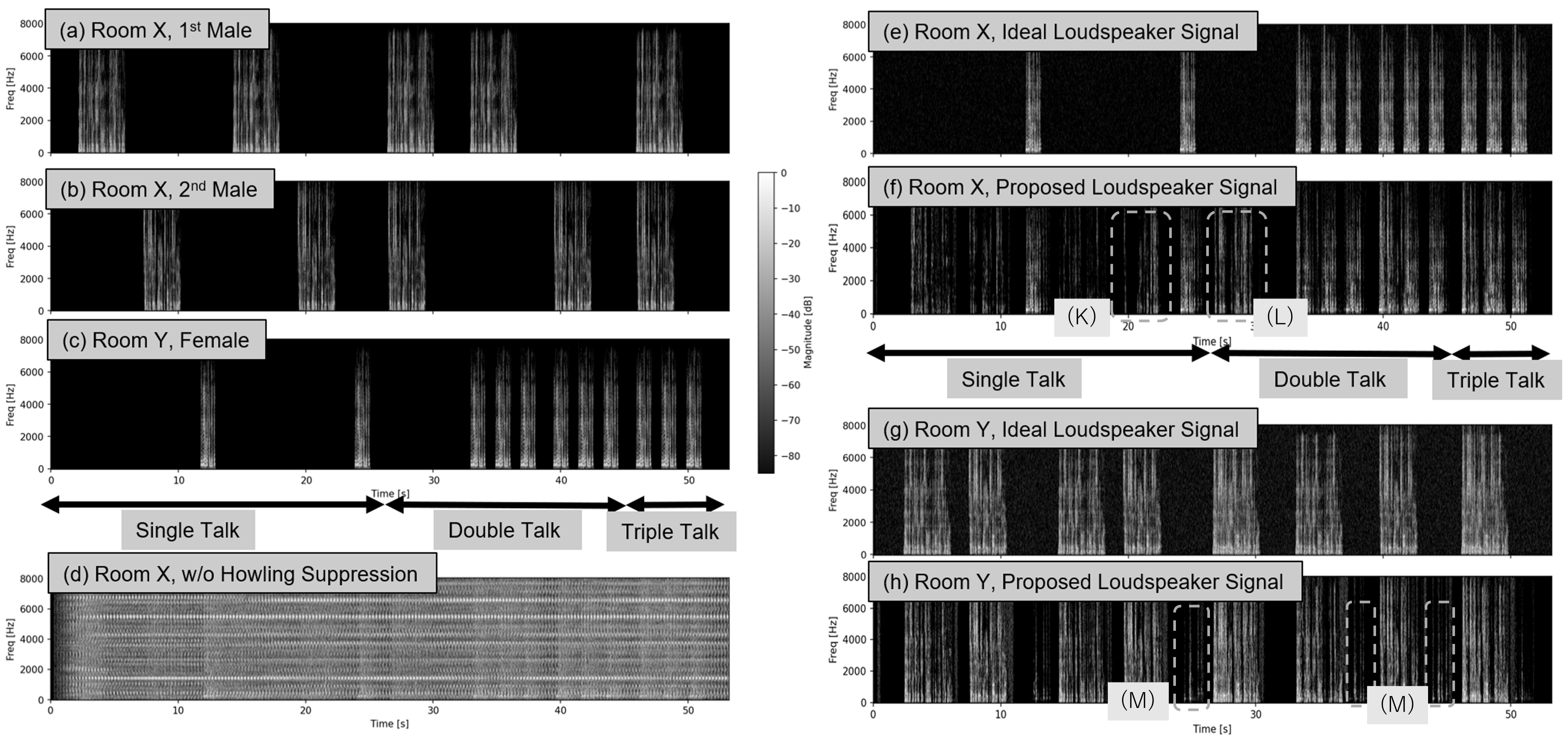}
\caption{Spectrograms from the verification simulation.
(a)--(c)~Talker source signals.
(d)~Howling without the proposed control.
(e),(g)~Ideal loudspeaker signals (face-to-face reference) in Rooms~A and~B.
(f),(h)~Loudspeaker signals with the proposed control in Rooms~A and~B.
Howling arises readily from terminal coupling in Room~A~(d), but is suppressed in~(f) and~(h).
Identification errors cause false passes at (K)--(M) and 
over-muting that thins the spectrograms, 
indicating that mute-only control is insufficient under double-talk.}
\label{SimResults202607}
\end{center}
\end{figure*}

\section{Expected Challenges}
Extraction and buffering are relatively easy;
identification and playback control dominate performance.
Placement, coexistence with the existing call pipeline,
and training/evaluation design are additional barriers.

\subsection{Identification}
Identification must be accurate and low-latency.
Extra delay beyond a few tens of milliseconds hurts interactivity,
so long-buffer high-accuracy matching is unattractive.
Short fragments lack features and raise false-pass errors
(same object judged different) and over-muting
(different objects judged the same).
False-pass errors seed howling; over-muting hurts intelligibility,
so the trade-off must be designed explicitly.

Spectra under comparison are strongly deformed from the original:
noise and interference, room reverberation, codecs~\cite{Brandenburg1991},
noise/echo suppressors~\cite{AudioProc2023Gael,Hoshuyama2006,Hoshuyama2012,MSAEC2008},
dynamic range control, fragmentation by user mute,
sampling-rate and clock mismatch~\cite{MS2022},
and time warping from packet loss or jitter compensation.
Large near/far talker level differences and continuous television or background music in the room
also complicate object boundaries and identity decisions.
Rephrasing, short backchannels, and repeated music phrases raise identification-error risk
even when utterances differ.

Candidate features include Mel-frequency cepstral coefficients~\cite{MFCC} and audio fingerprints~\cite{Cano2005Review,Foote1997,Haitsma2002,Wang2003}.
Short-segment matching is preferred to limit delay.
Fingerprints are robust but need redesign under ultra-low-latency and low-compute call constraints.
Longer buffers improve coverage but increase memory, computation, and accidental identification errors.
These issues resemble those in source separation and speech enhancement;
non-negative matrix factorization
and deep learning with deformation-robust representations are promising~\cite{Smaragdis2007,Kong2020},
yet inference delay, on-device compute, and generalization remain constraints.

\subsection{Playback Control}
Under simultaneous speech, hard mute is insufficient;
separation-like control that retains only needed objects is desirable.
Hard mute suppresses howling well but makes identification errors audible as dropouts.
Soft gain or frequency-selective mute can ease quality loss,
but residual leakage must not re-form a loop.
When identification is imperfect, context
(e.g., whether audio is from a TV in the same room, or whether the terminal is transmitting)
affects the mute decision.
Processing order relative to existing AEC, noise suppression, and codecs,
and the choice between joint AEC--NS~\cite{JointAEC-NS2019}
and task splitting~\cite{TaskSplit2022}, are open.
Placing the proposed control before or after AEC changes how references are deformed
and how hard identification becomes.

\subsection{Placement and System Implementation}
Processing may run on terminals, on the server, or both.
Terminal-side processing avoids extra uplink bandwidth and favors low delay,
but each terminal observes limited signals.
With at most one co-located terminal that does not implement the proposed control, howling can be prevented in principle,
but loops already formed elsewhere, or loops among unimplemented terminals, cannot be handled.
Server-side aggregation can improve identification but raises uplink delay, privacy, and load.
Under partial deployment, safe-side mute may over-mute and make calls impractical.
Thresholds and buffer lengths also depend on room, device, and codec conditions,
so stable operation without site-specific tuning is difficult.

\subsection{Training and Evaluation}
Public data that jointly cover network deformation, in-device nonlinear processing,
and playback control are scarce.
Conventional AEC benchmarks~\cite{AECchallenge2021} focus on intra-device echo
and do not cover howling via inter-terminal network paths.
Subjective quality and intelligibility assessment is also hard~\cite{Personalize2022Eskimez,intelligibility2010};
metrics must jointly capture howling suppression and over-mute degradation.
For supervised learning, defining identity labels
(how much time shift and deformation still count as the same object) is itself a research problem.
For privacy, storing fingerprints or embeddings rather than raw waveforms is preferable.

\section{Relation to Other Technologies}
The proposed approach extends voice-switched half-duplex~\cite{Busala1960,Haensler1992};
near-term use is echo control and howling suppression~\cite{Waterschoot2011}.
As half-duplex evolved into full-duplex via AEC,
adaptive processing, source separation, and deep learning can further improve performance.

In using reference signals, it resembles adaptive noise cancellation~\cite{Widrow1985}
and crosstalk cancellation~\cite{Sugiyama2008},
and can be seen as extending multi-channel AEC that is cast as source separation~\cite{Zhang2021}
to settings that include mute and communication paths.
It may also serve as a howling canceller for public-address systems,
but repeated sounds such as music raise identification-error risk.
Unlike personalized speech enhancement~\cite{Personalize2022Eskimez},
no speaker enrollment is required.
End-to-end learning that unifies identification and playback control is conceivable:
high-accuracy, delay-tolerant systems can provide data to fine-tune low-latency models.

\section{Verification Simulation}
To demonstrate the feasibility of sound-object-based echo control and to illustrate practical issues, a verification simulation is conducted
in the two-room, three-terminal setup of Fig.~\ref{EcSystemNwProp}.
The implementation is an initial gate based on cosine similarity of magnitude spectra,
not a final sound-object identifier.
Parameters were set empirically so that howling suppression is clear
while false-pass errors and over-muting under reverberation and double-talk remain observable.

\subsection{Verification Implementation}
Weighted overlap-add analysis is used at 16\,kHz with frame length 256 samples (16\,ms)
and hop 128 samples (8\,ms).
Utterance onset is declared when frame magnitude-spectrum energy exceeds a threshold;
termination follows four hops (32\,ms) below threshold, including two preceding hops (16\,ms).
This simplifies energy-based endpointing~\cite{Rabiner1975} with hangover~\cite{Sohn1999}.
Completed microphone-side objects are kept for 2.0\,s.
To limit decision delay, non-overlapping micro-objects of about 96\,ms are formed,
and candidates from the most recent 0.4\,s (0.35\,s for transmit-side references) are compared.
Receive-side loudspeaker control matches the latest six frames (48\,ms) every hop.

Identification uses cosine similarity of magnitude-spectrum sequences~\cite{Foote1997,Haitsma2002}.
Integer lags of $\pm 4$ hops ($\pm 32$\,ms) absorb propagation and processing delay;
pairs overlapping by fewer than three frames (24\,ms) are discarded.
On the receive side, pass gain 1.0 is used only if similarity $\le 0.66$; otherwise mute gain 0,
also when candidates are missing (safe-side).
On the transmit microphone side, mute if similarity $\ge 0.68$, else pass.
Gains are exponentially smoothed per hop (smoothing factor 0.88).
A1 uses both receive mute and transmit mute; A2 uses transmit mute only;
automatic mute on B1 is disabled.

\subsection{Call Chain and Talker Scenario}
The simulation jointly includes (1) same-room crosstalk,
(2) inter-room hands-free loudspeaker-to-microphone feedback, and
(3) moderate reverberation ($T_{60}=500$\,ms).
Background noise is about $-50$\,dBFS, and the inter-room delay is 200\,ms.
Each terminal cascades a frequency-domain partitioned AEC~\cite{Soo1990,Kellermann1988}
with double-talk-robust coefficient smoothing~\cite{Hoshuyama2008},
nonlinear residual echo suppression~\cite{Hoshuyama2006,Hoshuyama2012},
spectral subtraction~\cite{Martin2001},
and an Opus/CELT-like coding distortion model~\cite{Valin2012}.

Talkers are two scheduled male waveforms in Room~A and one female in Room~B,
covering single-talk and double- and triple-talker overlap.
Fig.~\ref{SimResults202607}(a)--(c) show the sources;
(d) shows howling about 2--3\,s after call start without the proposed control.

\subsection{Results}
Ideal loudspeaker signals as in face-to-face conversation 
are shown in Fig.~\ref{SimResults202607}(e) and~(g),
and the controlled signals in~(f) and~(h), respectively.
After AEC convergence (about 13\,s), single-talk identification errors are relatively few
and howling suppression is stable.
Double-talk and triple-talk (three-talker overlap) make similarity decisions ambiguous and control harder,
yet sustained howling is still suppressed.
Momentary echo from identification errors is quickly muted,
but over-muting fragments desired speech and reduces intelligibility.
The howling-suppression--quality trade-off remains a key topic for improvement.

\section{Conclusion}
Sound-object-based echo control is proposed
for suppressing howling caused by complicated acoustic paths
in multi-terminal voice calls, and the associated challenges are discussed.
The proposed approach shifts acoustic echo control from path estimation to sound object identification.
Default mute with pass/playback only when non-identity is established
breaks unintended echo loops and suppresses howling.
In a magnitude-spectrum similarity simulation, howling was suppressed,
but speech quality still suffered from over-muting.
Improving identification and playback control, especially with deep learning, is future work.

\subsection*{Acknowledgment}
Generative AI tools were used to assist with English writing and with parts of the simulation software development.
The author takes full responsibility for the manuscript and the code.


\begin{thebibliography}{99}

\bibitem{Sondhi1967}
M. M. Sondhi, ``An adaptive echo canceller,''
\textit{Bell Syst. Tech. J.}, vol. 46, no. 3, pp. 497--511, 1967.

\bibitem{Bennesty2001}
J. Benesty, T. Gansler, D. R. Morgan, M. M. Sondhi, and S. L. Gay,
\textit{Advances in Network and Acoustic Echo Cancellation}.
Berlin, Germany: Springer, 2001.

\bibitem{Enzner2014}
G. Enzner, H. Buchner, A. Favrot, and F. Kuech,
``Acoustic echo control,''
in \textit{Academic Press Library in Signal Processing}, vol. 4.
Oxford, U.K.: Elsevier, pp. 807--877, 2014.

\bibitem{Haensler2005}
E. H{\"a}nsler and G. Schmidt,
\textit{Acoustic Echo and Noise Control: A Practical Approach}.
Hoboken, NJ, USA: Wiley, 2005.

\bibitem{NeuralKalman2024}
E. Seidel, G. Enzner, P. Mowlaee, and T. Fingscheidt,
``Neural Kalman filters for acoustic echo cancellation:
Comparison of deep neural network-based extensions,''
\textit{IEEE Signal Process. Mag.}, vol. 41, no. 6, pp. 24--38, Nov. 2024.

\bibitem{Marini2024}
A. Marini, ``How we built Meet's Adaptive Audio feature,''
Google Blog, Nov. 1, 2024. [Online]. Available:
\url{https://blog.google/products-and-platforms/products/workspace/adaptive-audio-google-meet/}

\bibitem{Busala1960}
A. Busala,
``Fundamental considerations in the design of a voice-switched speakerphone,''
\textit{Bell Syst. Tech. J.}, vol. 39, no. 2, pp. 265--294, Mar. 1960.

\bibitem{Haensler1992}
E. H{\"a}nsler,
``The hands-free telephone problem: An annotated bibliography,''
\textit{Signal Process.}, vol. 27, no. 3, pp. 259--271, Jun. 1992.

\bibitem{Brandenburg1991}
K. Brandenburg and J. D. Johnston,
``Second generation perceptual audio coding: The hybrid coder,''
in \textit{Proc. AES 88th Convention}, Preprint 2937, Mar. 1990.

\bibitem{AudioProc2023Gael}
G. Richard, P. Smaragdis, S. Gannot, P. A. Naylor, S. Makino,
W. Kellermann, and A. Sugiyama,
``Audio signal processing in the 21st century:
The important outcomes of the past 25 years,''
\textit{IEEE Signal Process. Mag.}, vol. 40, no. 5, pp. 12--26, Jul. 2023.

\bibitem{Hoshuyama2006}
O. Hoshuyama and A. Sugiyama,
``An acoustic echo suppressor based on a frequency-domain model
of highly nonlinear residual echo,''
in \textit{Proc. IEEE Int. Conf. Acoust., Speech, Signal Process. (ICASSP)},
vol.~5, pp.~269--272, May 2006.

\bibitem{Hoshuyama2012}
O. Hoshuyama,
``An update algorithm for frequency-domain correlation model
in a nonlinear echo suppressor,''
in \textit{Proc. Int. Workshop Acoust. Echo Noise Control (IWAENC)},
Sep. 2012.

\bibitem{MSAEC2008}
D. A. Bendersky, J. W. Stokes, and H. S. Malvar,
``Nonlinear residual acoustic echo suppression for high levels of harmonic distortion,''
in \textit{Proc. IEEE Int. Conf. Acoust., Speech, Signal Process. (ICASSP)},
pp. 261--264, Mar./Apr. 2008.

\bibitem{MS2022}
E. Indenbom, N.-C. Ristea, A. Saabas, T. P{\"a}rnamaa, and J. Gu{\v{z}}vin,
``Deep model with built-in cross-attention alignment for acoustic echo cancellation,''
arXiv:2208.11308, 2022.

\bibitem{MFCC}
Z. K. Abdul and A. K. Al-Talabani,
``Mel frequency cepstral coefficient and its applications: A review,''
\textit{IEEE Access}, vol. 10, pp. 122136--122158, 2022.

\bibitem{Cano2005Review}
P. Cano, E. Batlle, T. Kalker, and J. Haitsma,
``A review of audio fingerprinting,''
\textit{J. VLSI Signal Process. Syst.}, vol. 41, no. 3, pp. 271--284, Nov. 2005.

\bibitem{Foote1997}
J. Foote,
``Content-based retrieval of music and audio,''
in \textit{Multimedia Storage and Archiving Systems II, Proc. SPIE},
vol. 3229, pp. 138--147, 1997.

\bibitem{Haitsma2002}
J. Haitsma and T. Kalker,
``A highly robust audio fingerprinting system,''
in \textit{Proc. Int. Conf. Music Inf. Retrieval (ISMIR)},
pp. 107--115, Oct. 2002.

\bibitem{Wang2003}
A. L. Wang,
``An industrial-strength audio search algorithm,''
in \textit{Proc. Int. Conf. Music Inf. Retrieval (ISMIR)},
pp. 7--13, Oct. 2003.

\bibitem{Smaragdis2007}
P. Smaragdis, B. Raj, and M. Shashanka,
``Supervised and semi-supervised separation of sounds from single-channel mixtures,''
in \textit{Proc. Int. Conf. Independent Component Analysis and Signal Separation (ICA)},
pp.~414--421, 
Sep. 2007.

\bibitem{Kong2020}
Q. Kong, Y. Cao, T. Iqbal, Y. Wang, W. Wang, and M. D. Plumbley,
``PANNs: Large-scale pretrained audio neural networks for audio pattern recognition,''
\textit{IEEE/ACM Trans. Audio, Speech, Lang. Process.},
vol. 28, pp. 2880--2894, 2020.

\bibitem{JointAEC-NS2019}
H. Zhang, K. Tan, and D. Wang,
``Deep learning for joint acoustic echo and noise cancellation with nonlinear distortions,''
in \textit{Proc. Interspeech}, pp.~4255--4259, Sep. 2019.

\bibitem{TaskSplit2022}
S. Braun and M. L. Valero,
``Task splitting for DNN-based acoustic echo and noise removal,''
in \textit{Proc. Int. Workshop Acoust. Signal Enhancement (IWAENC)},
pp.~1--5, Sep. 2022.

\bibitem{AECchallenge2021}
R. Cutler \textit{et al.},
``Acoustic echo cancellation challenge,''
in \textit{Proc. Interspeech}, pp. 4274--4278, Aug. 2021.

\bibitem{Personalize2022Eskimez}
S. E. Eskimez, T. Yoshioka, H. Wang, X. Wang, Z. Chen, and X. Huang,
``Personalized speech enhancement: New models and comprehensive evaluation,''
in \textit{Proc. IEEE Int. Conf. Acoust., Speech, Signal Process. (ICASSP)},
pp. 356--360, May 2022.

\bibitem{intelligibility2010}
P. C. Loizou and G. Kim,
``Reasons why current speech-enhancement algorithms do not improve speech intelligibility
and suggested solutions,''
\textit{IEEE Trans. Audio, Speech, Lang. Process.},
vol. 19, no. 1, pp. 47--56, Jan. 2011.

\bibitem{Waterschoot2011}
T. van Waterschoot and M. Moonen,
``Fifty years of acoustic feedback control: State of the art and future challenges,''
\textit{Proc. IEEE}, vol. 99, no. 2, pp. 288--327, Feb. 2011.

\bibitem{Widrow1985}
B. Widrow and S. D. Stearns,
\textit{Adaptive Signal Processing}.
Englewood Cliffs, NJ, USA: Prentice-Hall, 1985.

\bibitem{Sugiyama2008}
A. Sugiyama,
``Low distortion noise cancellers---revival of a classical technique,''
in \textit{Speech and Audio Processing in Adverse Environments},
E. H{\"a}nsler and G. Schmidt, Eds.
Berlin, Germany: Springer, pp. 229--264, 2008.

\bibitem{Zhang2021}
H. Zhang and D. Wang,
``A deep learning approach to multi-channel and multi-microphone acoustic echo cancellation,''
in \textit{Proc. Interspeech}, pp. 2461--2465, Aug. 2021.

\bibitem{Rabiner1975}
L. R. Rabiner and M. R. Sambur,
``An algorithm for determining the endpoints of isolated utterances,''
\textit{Bell Syst. Tech. J.}, vol. 54, no. 2, pp. 297--315, Feb. 1975.

\bibitem{Sohn1999}
J. Sohn, N. S. Kim, and W. Sung,
``A statistical model-based voice activity detection,''
\textit{IEEE Signal Process. Lett.}, vol. 6, no. 1, pp. 1--3, Jan. 1999.

\bibitem{Soo1990}
J.-S. Soo and K. K. Pang,
``Multidelay block frequency domain adaptive filter,''
\textit{IEEE Trans. Acoust., Speech, Signal Process.},
vol. 38, no. 2, pp. 373--376, Feb. 1990.

\bibitem{Kellermann1988}
W. Kellermann,
``Analysis and design of multirate systems for cancellation of acoustical echoes,''
in \textit{Proc. IEEE Int. Conf. Acoust., Speech, Signal Process. (ICASSP)},
pp. 2570--2573, Apr. 1988.

\bibitem{Hoshuyama2008}
O. Hoshuyama,
``An echo canceller using smoothed-coefficient filter with adaptive time constant
controlled by high-pass errors,''
in \textit{Proc. Int. Workshop Acoust. Echo Noise Control (IWAENC)},
Paper ID 9015, Sep. 2008.

\bibitem{Martin2001}
R. Martin,
``Noise power spectral density estimation based on optimal smoothing and minimum statistics,''
\textit{IEEE Trans. Speech Audio Process.},
vol. 9, no. 5, pp. 504--512, Jul. 2001.

\bibitem{Valin2012}
J.-M. Valin, K. Vos, and T. B. Terriberry,
``Definition of the Opus audio codec,''
IETF RFC~6716, Sep. 2012. [Online]. Available:
\url{https://www.rfc-editor.org/rfc/rfc6716}

\end{thebibliography}
\end{document}